\documentclass[prd,aps,showpacs,nofootinbib,floats,floatfix,superscriptaddress,twocolumn,letterpaper]{revtex4}

\usepackage{psfrag}
\usepackage[dvips]{graphicx}
\usepackage{amsmath}
\usepackage{amsfonts}
\usepackage{color}
\usepackage{dcolumn}
\usepackage{bm}
\usepackage{amsmath}

\def\laq{\raise 0.4ex\hbox{$<$}\kern -0.8em\lower 0.62ex\hbox{$\sim$}}
\def\gaq{\raise 0.4ex\hbox{$>$}\kern -0.7em\lower 0.62ex\hbox{$\sim$}}

\newcommand{\beq}{\begin{equation}}
\newcommand{\eeq}{\end{equation}}
\newcommand{\bea}{\begin{eqnarray}}
\newcommand{\eea}{\end{eqnarray}}
\newcommand{\ba}{\begin{array}}
\newcommand{\ea}{\end{array}}

\newcommand{\Maryland}{\affiliation{Maryland Center for Fundamental
    Physics, Department of Physics, University of Maryland, College
    Park, MD 20742}}

\newcommand{\Cornell}{\affiliation{Center for Radiophysics and Space
    Research, Cornell University, Ithaca, New York, 14853}}

\newcommand{\LSU}{\affiliation{Department of Physics and Astronomy,
    Louisiana State University, Baton Rouge, LA 70803-4001}}

\begin{document}

\title{Estimating the final spin of a binary black hole coalescence}

\author{Alessandra Buonanno} \Maryland
\author{Lawrence E. Kidder} \Cornell
\author{Luis Lehner} \LSU

\begin{abstract}
We present a straightforward approach for estimating the final black
hole spin of a binary black hole coalescence with arbitrary initial
masses and spins. Making some simple assumptions, we estimate the
final angular momentum to be the sum of the individual spins plus the
orbital angular momentum of a test particle orbiting at the last
stable orbit around a Kerr black hole with a spin parameter of the
final black hole. The formula we obtain is able to reproduce with
reasonable accuracy the results from available numerical simulations,
but, more importantly, it can be used to investigate what
configurations might give rise to interesting dynamics.  In
particular, we discuss scenarios which might give rise to a ``flip''
in the direction of the total angular momentum of the system. By
studying the dependence of the final spin upon the mass ratio and
initial spins we find that our simple approach suggests that it is not
possible to spin-up a black hole to extremal values through merger
scenarios irrespective of the mass ratio of the objects involved.
\end{abstract}


\maketitle

\section{Introduction}

Our understanding of the dynamics of a binary black hole system has
advanced at an impressive pace in the last several years as numerical
simulations have provided the missing piece for describing the
complete dynamics of the system, starting from the late inspiral, and
continuing through the merger and ringdown of the final black hole.
While these simulations have been exploring the parameter space of
possible configurations of initial masses and spins of the black
holes, the process is necessarily slow due to the time consuming
nature of the simulations.  This has led to a number of recent works
aimed at producing recipes that allow for predicting, to a certain
extent, what can be expected as the final state of the merged black
hole from a given initial configuration.  This is important, not only
in order to decide which situations might give rise to the most
interesting dynamical behavior, but also for the possibility of
exploiting this information within a simulation. Two examples of this
are (1) in the extraction of waveforms, where choosing an adequate
background can reduce the errors in determining observable effects,
and (2)  in helping to provide an {\it analytic} description of 
inspiral-merger-ringdown to be used in the construction of templates.

To date efforts to produce such estimates have progressed on several
fronts.  The effective-one-body (EOB) approach~\cite{BD1,DJS}, which
maps the two-body problem to the dynamics of a test-particle in a
suitably defined spacetime, provided estimates for the final black
hole spin~\cite{BD2,TD,BCD} that turn out to agree with current
numerical simulations to within $\sim 10\%$.  More recently, by
combining the EOB approach with test-mass limit predictions for the
energy released during the merger and ringdown phases, Ref.~\cite{DN} has
refined these predictions for the non-spinning case to obtain final
spins that agree to within $\sim 2\%$.  On another front, several
fitting formulae for the final black hole spin have recently appeared
in the literature. These formulae are based on the results of
available numerical simulations along with a judicious exploitation of
the natural symmetries in the problem and/or expectations from the
test-mass limit (see e.g. Refs.~\cite{{loustofit, Berti:2007fi, 
EOBp4PN,Rezzolla:2007xa, Marronetti:2007wz, Boyle:2007sz}}).

In this work we follow a different route based on first principles,
although implicitly using numerical results along with intuitive
arguments that during the merger and ringdown phases, the mass and
angular momentum of the system are roughly conserved. Beyond this
assumption, however, and in contrast to the approaches mentioned
above, we make no explicit use of results from post-Newtonian
approximations or numerical simulations.  Our straightforward approach
is based on simple estimates for the quasi-adiabatic inspiral, coupled
with standard results for the angular momentum of a particle in a Kerr
black hole spacetime corresponding to the {\em final} black hole.  We
obtain a closed form expression for the final spin of the black hole
for arbitrary mass ratios and spins. We find that predictions from
this simple-minded expression give results that agree reasonably well
with the numerical simulations.  This is yet further evidence of the
rather simple behavior underlying the dynamics of binary black hole
spacetimes.

In spirit, our work is similar to that of Ref.~\cite{Hughes:2002ei}, who
used a point particle approximation on a Kerr background to
estimate the mass and spin of the final black hole. Our approach,
however, contains an important difference with that
of Ref.~\cite{Hughes:2002ei}, which allows us to make reasonable
predictions even for comparable mass systems.

Our approach and assumptions are described in
Sec.~\ref{sec:phen-appr}, and we present a simple expression that
estimates the spin of the final black hole. In
Sec.~\ref{sec:representative-cases} we illustrate our results for
several interesting scenarios, including spinning, precessing 
black hole binary systems, and compare with numerical
simulations. We conclude in Sec.~\ref{sec:final-comments} with
final comments.

\section{The approach}
\label{sec:phen-appr}

We consider two widely separated black holes that can be well approximated
by two Kerr black holes with masses and angular momentum
parameters $(m_i, a_i)$.  For simplicity we will first
restrict our discussion to scenarios
where the orbit stays within a plane (which we will refer to as the
equatorial plane). In this case the orbital angular momentum and
individual spins are orthogonal to the equatorial plane.  Our goal is
to obtain the angular momentum of the final black hole in terms of the
initial configuration of the system by a phenomenological approach
rather than by evolving the system numerically. This will aid in
finding particularly interesting cases that can be explored with
numerical simulations.

Achieving this goal certainly requires some compromises as the system
undergoes a long dynamical process that comprises several stages:
inspiral, merger and ringdown. While an accurate description of the
whole process requires following the system completely, a back of the
envelope estimate can be obtained by exploiting the fact that: (i)
during the inspiral phase, the system evolves quasi-adiabatically, and
(ii) during the merger and ringdown phases the total mass and angular
momentum of the system change by only a small amount. These
observations are based on intuitive arguments but are strongly backed
by post-Newtonian and perturbative calculations for the inspiral and
ringdown, and by numerical simulations for the merger.

We obtain a simple expression, based on first principles, that can be
employed to predict what the angular momentum parameter of the final
black hole will be. This expression is obtained naturally when,
inspired by the observations above, the following assumptions are
made:
\begin{itemize}
\item To first order, the mass of the system is conserved. Thus, the
  final black hole will have total mass $M=m_1+m_2$. During the whole
  process the total radiated energy remains small ($M_{\rm radiated}
  \,\laq\, 10\%\,M_{\rm initial}$); thus this assumption is
  justified to the level of approximation we seek.
\item The magnitude of the individual spins of the black holes will
  remain constant. Since both spin-spin and spin-orbit couplings are
  small, and radiation falling into the black holes affects the spins
  by a small amount, this is a safe assumption.  Therefore the
  contribution to the final total angular momentum due to the
  individual black holes spins will be determined by the initial
  spins.
\item The system radiates much of its angular momentum in the long
  inspiral stage until it reaches the innermost stable circular orbit
  (ISCO), at which point the dynamics quickly leads to the merger of
  the black holes. During the merger the radiation of energy and angular
  momentum with respect to the mass and angular momentum of the system
  is small.  Thus, to estimate the contribution of the orbital angular
  momentum to the final angular momentum of the black hole, we adopt
  the orbital angular momentum of a test-particle orbiting at the ISCO
  of a Kerr black hole with a spin parameter of the {\em final} black
  hole. While adopting this value makes strict sense in the extreme
  mass ratio case, we will see that it leads to predictions that agree
  quite well with the results of numerical simulations.  That the
  point-particle approximation is able to capture key aspects of the
  two-body dynamics has also been observed in
  Refs.~\cite{BD2,BCP,DN,EOBp4PN} when comparing with results in
  Ref.~\cite{Davis}, in Refs.~\cite{CLA,lazarus} when comparing 
  the close-limit approximation to numerical simulations, and
  more recently in studies of unstable circular orbits in black hole
  mergers~\cite{Pretorius:2007jn}.
\end{itemize}
Bringing all these assumptions together, we may express the angular
momentum parameter of the final black hole $a_f$ as,
\begin{equation}
\frac{a_f}{M} = \frac{L_{\rm orb}}{M^2}(r_{\rm ISCO},a_f)
+ \frac{m_1 a_1}{M^2} + \frac{m_2 a_2}{M^2}\,,
\label{af}
\end{equation}
where $L_{\rm orb}$ is the orbital angular momentum of a particle at the
ISCO of a Kerr black hole (with spin parameter $a_f$).  

Note that our assumptions differ in several ways from those of
Ref.~\cite{Hughes:2002ei}: (1) we keep the mass of the system
constant, while Ref.~\cite{Hughes:2002ei} adds a contribution to the
final mass from the energy at the ISCO; (2) we keep the contributions
from the spins of both bodies, while Ref.~\cite{Hughes:2002ei}
neglects the spin of the smaller black hole; and (3) we use the
orbital angular momentum of the ISCO for a Kerr black hole with the
{\em final} spin, while Ref.~\cite{Hughes:2002ei} uses the initial
spin of the larger black hole.  For extreme mass ratio cases, both
approaches would give essentially the same answer; however our
approach can be applied to general mass ratios and accounts for 
the orbital and both individal spin angular momenta when obtaining the
final black hole spin.

We can re-express our formula for $a_f$ in a more convenient form as
\bea
\frac{a_f}{M} &=& \frac{L_{\rm orb}}{M^2} (r_{\rm ISCO},a_f) + \frac{\chi_1}{4}
     {\left(1+\sqrt{1-4\nu}\right)^2} \nonumber \\
 && + \frac{\chi_2}{4}
     {\left(1-\sqrt{1-4\nu}\right)^2}\,, \label{mastereqn}
\eea
where $\chi_i=a_i/m_i$ ($\in [-1,1]$) and $\nu = (m_1 m_2)/M^2$ ($\in
[0,1/4]$), and without loss of generality it is assumed that $m_1 \,\gaq\,
m_2$.  This equation provides a way to obtain $a_f$ given $m_i$ and
$\chi_i$.  Since the expression for $L_{\rm orb}$ at the ISCO is known
in closed form for equatorial orbits we concentrate first on such
cases. This will allow us to investigate the viability
of Eq.~(\ref{mastereqn}) by comparing it to results obtained
with numerical simulations.

Adopting the expression for equatorial orbits in Ref.~\cite{bardeenpressteukolsky} one deduces,
\begin{equation} \label{eq:KerrL}
\frac{L_{\rm orb}}{M^2} = \pm \frac{\nu(r^{2}\mp2 a_f M^{1/2} r^{1/2} + a_f^2)}
{M^{1/2} r^{3/4}(r^{3/2} - 3 M r^{1/2} \pm 2 a_f M^{1/2} )^{1/2}}\,,
\end{equation}
where the upper signs correspond to prograde orbits and the lower signs to
retrograde orbits.  This function is to be evaluated at $r=r_{\rm ISCO}$ with
\beq
r_{\rm ISCO} = M \left\{ 3 + Z_2 \mp \left[ (3-Z_1)(3 + Z_1 + 2 Z_2)
\right]^{1/2} \right\}\,, 
\eeq
where
\bea
Z_1 &\equiv& 1 + \left (1-\frac{a_f^2}{M^2} \right )^{1/3}
\left [ \left (1+\frac{a_f}{M} \right )^{1/3}+ 
\left (1-\frac{a_f}{M} \right )^{1/3}\right ]\,,
\nonumber \\ 
Z_2 &\equiv& \left ( 3 \frac{a_f^2}{M^2} + Z_1^2 \right
)^{1/2}\,. \nonumber
\eea
The use of the prograde or retrograde case depends on whether the
final spin is aligned or anti-aligned with the initial orbital angular
momentum.  Indeed one particularly interesting application of the above
expression is to understand the direction of the final spin as a
function of initial spins and the mass ratio.

\section{Representative cases}
\label{sec:representative-cases}

\begin{figure}
\begin{center}
\includegraphics[width=8.cm,angle=-90]{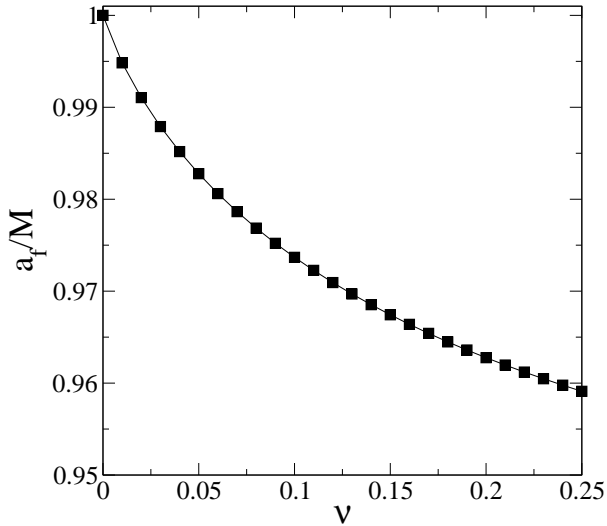}
\caption{\label{Fig:maxspin} The final spin $a_f/M$ vs $\nu = m_1
  m_2/M^2$ for initial black holes with extreme spin parameters
  $\chi=1$. Clearly, as the mass-ratio approaches $\nu=1/4$ (equal-mass 
  case) the final expected spin parameter decreases.}
\end{center}
\end{figure}

We now concentrate on several representative cases that explore
different interesting scenarios, and make contact with available
numerical results.

\subsection{Equal spin case}
\label{sec:equal-spin-case}

A simple case that can be compared with existing simulations is for
equal spins (i.e.  $\chi_1 = \chi_2 = \chi$). In this case
equation (\ref{mastereqn}) reduces to 
\begin{equation} 
\label{equalspins}
\frac{a_f}{M} =  \frac{L_{\rm orb}}{M^2} + (1-2\nu) \chi\,.
\end{equation}
This equation allows us to determine the value of $a_f$ as a function
of $\nu$ and $\chi$ and answer specific questions. Figure
\ref{Fig:maxspin} illustrates the behavior of the final spin parameter
as a function of mass ratio when the individual spins of the initial
black holes are maximal.  The largest spin for the final black hole is
achieved for the extreme mass ratio case. This coincides with the
intuitive picture that a particle falling into an extreme black hole
will have negligible effect on the final spin, while a head-on
collision of equal-mass extreme Kerr black holes will give rise to a
final black hole with $a_f/M = 1/2$.

\begin{figure}
\begin{center}
\includegraphics[width=8.cm,angle=-90]{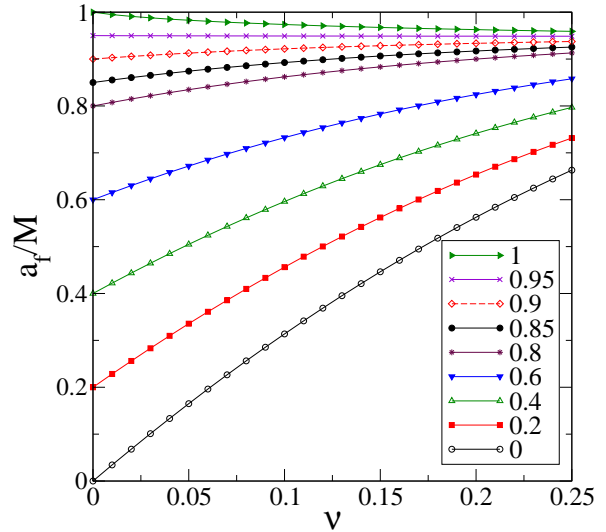}
\caption{\label{Fig:spinsweep} Final spin $a_f/M$ vs $\nu$ 
for initial black holes for several spin parameters
$\chi=0,0.2,0.4,0.6,0.8,0.85,0.9,0.95,1$.  For spins $\gaq\, 
0.948$ the maximum final spin is achieved as the extreme mass ratio
case is approached while for initial spins $\,\laq\, 0.948$ the
equal-mass case would give rise to the maximum final spin.}
\end{center}
\end{figure}

This behavior, however, varies when considering initial spins less
than maximal. For initially non-spinning black holes, intuitively the
final black hole will also be essentially non-spinning for the extreme
mass ratio case while it would have a non-trivial final spin for the
equal-mass case.  Figure \ref{Fig:spinsweep} illustrates the spin of
the final black hole as a function of the mass ratio for different
values of $\chi$ for spins that are aligned with the orbital angular
momentum.  We see that there is a critical value for the initial
spins, $\chi = 0.948$, above which, the maximum final spin will
increase as the mass ratio $q=m_1/m_2 \geq 1$ increases.  Below the
critical value, the final spin will increase as the equal-mass limit
is approached. Finally, at the critical value any merger will leave
the final spin essentially unchanged irrespective of the mass ratio.

\begin{figure}
\begin{center}
\includegraphics[width=8.cm,angle=-90]{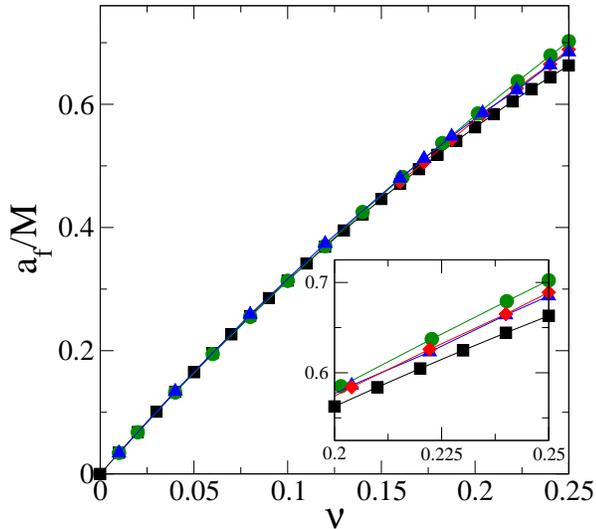}
\caption{\label{Fig:noindspin} Final spin $a_f/M$ vs $\nu$ for initial
  black holes with initially non-spinning black holes ($\chi=0$) as
  predicted by our simple model (filled squares), and as obtained
  numerically in Ref.~\cite{Berti:2007fi} (red diamonds), 
  by the EOB model combined with non-spinning test-particle predictions in
  Ref.~\cite{DN} (green circle) and by numerical relativity combined 
  with non-spinning test-particle expectations in 
  Ref.~\cite{EOBp4PN} (blue triangle). The largest final black hole
  spin is obtained in the equal-mass case.}
\end{center}
\end{figure}

The case of black holes which are initially non-spinning can be
compared directly with a number of simulations~\cite{Pret05, CLMZ06,
  BCCKM06, DHPS06, GSBHH07}.  For equal masses, the value predicted by
Eq.~(\ref{equalspins}) is $a_f/M=0.66$, which is quite close to the
value $a_f/M=0.68$ obtained by simulations of equal-mass, non-spinning
black holes. Figure \ref{Fig:noindspin} illustrates our predicted
values for the final spin as the mass ratio is varied from the extreme
mass ratio to the equal-mass case, along with the results from
 Ref.~\cite{DN, Berti:2007fi, EOBp4PN}.  Excellent agreement is found with
results from numerical simulations. Notice that the final black hole
with the highest spin occurs in the equal-mass case as expected since
the orbital angular momentum is maximized in that case.

\begin{figure}
\begin{center}
\includegraphics[width=8.cm,angle=-90]{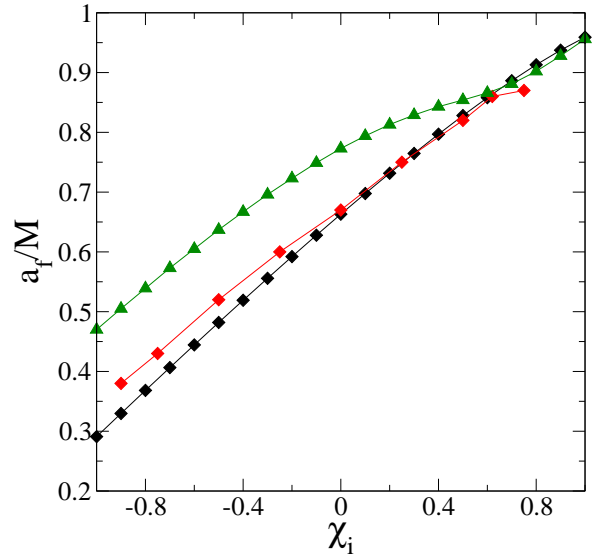}
\caption{\label{Fig:equalmass_spinvar} Final spin $a_f/M$ vs $\chi$
  for equal-mass black holes as predicted by our expression (filled
  squares), numerically obtained (filled diamonds) in
  Ref.~\cite{Marronetti:2007wz} and from the EOB model (filled triangle) 
   used in Ref.~\cite{BCD}, but without taking into account the 
   angular momentum released during merger-ringdown (thus, the 
  latter curve overestimates the results and should be considered 
   as an upper limit). Irrespective of the alignment or
  anti-alignment of the individual spins, the final black hole spin is
  aligned with the initial orbital angular momentum.}
\end{center}
\end{figure}

Another case that can be compared to simulations is for equal masses
where the initial spins are either aligned or anti-aligned with
respect to the orbital angular momentum. While simulations for close
to maximally spinning black holes are a challenge, robust results
exist for $\chi_i \in [-0.7,0.7]$.  Figure \ref{Fig:equalmass_spinvar}
shows the final value for the spin parameter predicted by
Eq.~(\ref{equalspins}) as $\chi$ is varied, along with the values
available from existing simulations. Clearly, a quite reasonable
agreement is found for $\chi_i \in [-0.7,0.7]$. Furthermore, we obtain
for the extreme cases $a_f({\chi=-1})=0.2909 M$ and
$a_f({\chi=1})=0.9591 M$, which are, respectively, $14\%$ and $0.01\%$
away from the values reported by the fit formulae employed in
Ref.~\cite{Marronetti:2007wz}.  Figure~\ref{Fig:equalmass_spinvar} also
shows results from the EOB model used in Ref.~\cite{BCD}, where the final
black hole spin is computed at the end of the EOB plunge, disregarding
spin-spin corrections and without taking into account the angular
momentum released during merger-ringdown~\footnote{Note that the
  (conservative) EOB Hamiltonian used in Ref.~\cite{BCD} differs from the
  one used in Ref.~\cite{TD}.  Whereas Ref.~\cite{BCD} adds spin effects to the
  (resummed) non-spinning EOB Hamiltonian, which is a deformation of
  the Schwarzschild spacetime, Ref.~\cite{TD} also includes spin effects in
  the resummation, obtaining an EOB Hamiltonian which is a deformation
  of a Kerr spacetime.}; thus, the values should be considered as an
upper limit~\footnote{For example in the non-spinning case ($\chi_i =
  0$ in Fig.~\ref{Fig:equalmass_spinvar}) by including test-mass limit
  predictions for the angular momentum released during the
  merger-ringdown phases, Ref.~\cite{DN} has reduced the difference from
  $\sim 10 \%$ to $\sim 2\%$.}.

\begin{figure}
\begin{center}
\includegraphics[width=7.2cm,angle=-90]{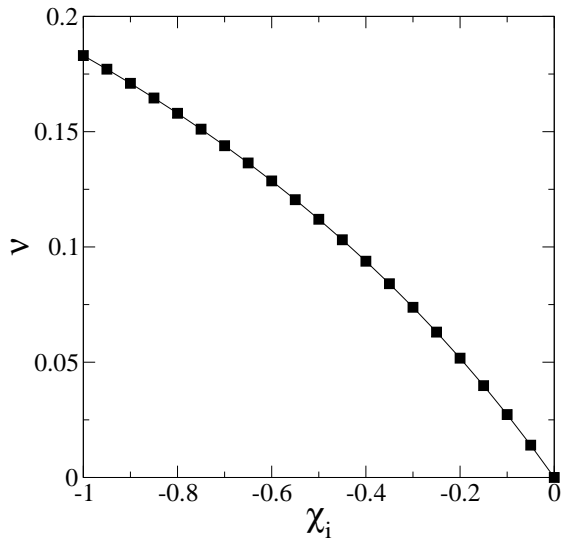}
\caption{\label{Fig:final0spin} Relation between the mass ratio $\nu$
  and the individual spins of the black holes to achieve a final
  non-spinning black hole. Only mass ratios $\nu \,\laq\, 0.183$
  ($q \,\gaq\, 3.15$)
  will allow for the final spin to be anti-aligned with the initial
  orbital angular momentum direction if the individual black holes
  spin fast enough.} 
\end{center}
\end{figure}

A final interesting example that we report here is where the
individual spins are anti-aligned with the {\it initial} orbital
angular momentum and the spin of the final black hole is zero.  This
occurs in our model when the orbital angular momentum remaining at the
ISCO is exactly counteracted by the individual black hole spins. This
scenario determines the border between cases in which the direction of
the final angular momentum (or final spin) is determined by the
initial orbital angular momentum, or by the direction of the initial
black hole spins. In the latter cases the direction of the total
angular momentum will ``flip'', and inertial frames will see the
direction they are dragged reverse as the system goes through the
merger. This should be reflected in the waveforms, which would likely
display an interesting behavior~\cite{ACST}.
Figure~\ref{Fig:final0spin} shows the mass ratios and initial value of
the individual anti-aligned spins required to yield a final
non-spinning black hole.  For a realizable situation of $a_i=0.8 m_i$
the mass ratio adopted should be $4:1$, while for maximally spinning
holes, we would predict a final spin of zero for the mass ratio $q
\simeq 3.15$. Note that this value is smaller than that predicted by
Ref.~\cite{Hughes:2002ei} of $q \simeq 4.23$.  By our arguments, any
system with mass ratio $q>6.78$ will undergo a flip of the total
angular momentum if the individual spins are equal and anti-aligned
with the initial orbital angular momentum and their spin parameters
obey $a_i/m_i \geq 1/2 $. Notice however that the orbital separation
at which this flip takes place also depends upon the mass ratio.  This
location can be estimated via simple Newtonian arguments.  A particle
of reduced mass $\mu=\nu M$ in a circular orbit about a central object
of mass $M$ has an associated orbital angular momentum given by,
\begin{equation}
L_{\rm approx} = \mu \sqrt{r M} \, ,
\end{equation}
Thus, the distance at which $L_{\rm approx}$ will be canceled by the
 contribution of the individual spins is determined by
\begin{equation}
\frac{L_{\rm approx}}{M^2} = (1-2\nu) |\chi|
\end{equation}
Thus, the following estimate 
\begin{equation}
\frac{r}{M} = \frac{(1-2\nu)^2}{\nu^2} |\chi|^2
\end{equation}
can be used to determine the approximate distance at which the flip
will occur.  Notice in the limit $\nu~\rightarrow~0$
($q~\rightarrow~\infty$), that $r~\rightarrow~\infty$ so the flip has
essentially taken place prior to any astrophysically interesting
initial configuration.  In such cases the final spin direction is
determined by the spin of the large black hole.

\subsection{Unequal spin case}

\begin{figure}
\begin{center}
\includegraphics[width=7.5cm,angle=-90]{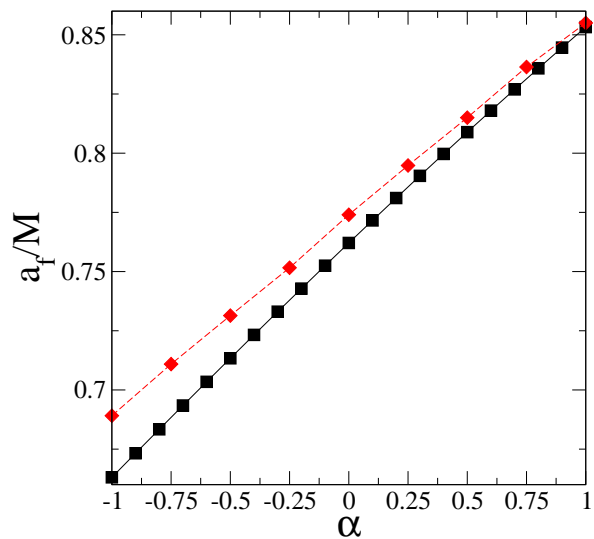}
\caption{\label{Fig:diffspin_eqmass} Final spin $a_f/M$ vs $\alpha$ for
  equal-mass black holes with $\chi_1=0.584$ and $\chi_2 \in
  [-0.584,0.584]$ as predicted by our simple model (filled squares)
  and as obtained numerically in Ref.~\cite{Rezzolla:2007xa} (red
  diamonds).}
\end{center}
\end{figure}

\begin{figure}
\begin{center}
\includegraphics[width=8.cm,angle=-90]{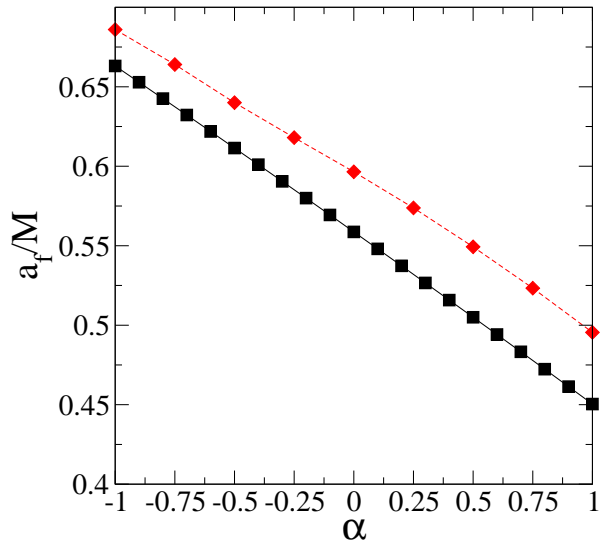}
\caption{\label{Fig:diffspin_eqmass_down} Final spin $a_f/M$ vs $\alpha$
  for equal-mass black holes with $\chi_1=-0.584$ and $\chi_2 \in
  [-0.584,0.584]$ as predicted by our simple model (filled squares)
  and as obtained numerically in Ref.~\cite{Rezzolla:2007xa} (red
  diamonds).}
\end{center}
\end{figure}

We now illustrate the case with equal masses, but
unequal spins. Setting $\chi_1 = \chi = \alpha \chi_2$ and $\nu =
1/4$ in Eq.~(\ref{mastereqn}) yields
\begin{equation}
\frac{a_f}{M} = \frac{L_{\it orb}}{M^2} + \frac{\chi}{4} (1 + \alpha)
\label{nonequalspins} \,.
\end{equation}
Figure~\ref{Fig:diffspin_eqmass} illustrates the value of the final
spin parameter for equal-mass black holes and $\chi=0.584$ and varying
$\alpha \in[-1,1]$, while Fig.~\ref{Fig:diffspin_eqmass_down}
illustrates the case with $\chi=-0.584$.  In both cases we compare
directly against the results of the simulations in
Ref.~\cite{Rezzolla:2007xa}. Our results differ at most by $8\%$ with
the reported results. We have also compared with simulations with mass
ratio $3:2$ and either $a_1/m_1=-0.194$ and $ a_2/m_2= 0.201$, or
$a_1/m_1=0.193$ and $a_2/m_2= -0.201$~\cite{Goddard}.  In these cases
our simple model predicts a final spin of $0.617 M$ and $0.671 M$,
respectively, while Ref.~\cite{recoil} obtains numerically a final
spin of $0.640 M$ and $0.704 M$.  Finally, for equal-masses and
$a_1/m_1=-0.198$ and $a_2/m_2= 0.198$, Ref.~\cite{Goddard,recoil}
obtains a final spin of $0.697 M$, whereas we predict $0.663 M $.

\subsection{Generic spin configurations}
\label{sec3C}

Until now our analysis has been restricted to cases where the
orbital plane does not change in time. However, we expect that 
the same arguments which lead us to Eq.~(\ref{af}) 
are applicable to more generic scenarios with precessing orbits 
and arbitrary directions for the individual spin. A key 
difference in generic cases is that the orbit at the ISCO will, in 
general, be inclined with respect to the final total angular momentum. 
For these cases the expression for the orbital 
contribution to the total angular momentum would require either the numerical
integration of generic geodesics in a Kerr spacetime or the use of the radial
potential for quasi-adiabatic spherical orbits~\cite{Chandra}. Alternatively,
one can make use of the fit formulae presented in Ref.~\cite{Hughes:2002ei}
 to express $\vec L_{orb}$ in analytic form.

The simplest possible extension of our method to more generic spin 
configurations can be formulated when the following 
assumptions (in addition to the 
assumptions we made in Sec.~\ref{sec:phen-appr}) are adopted:
\begin{itemize}
\item The following quantities are known
\[
\{m_1,m_2,{\vec S_1},{\vec S_2},{\hat L}_{\rm orb}\}
\]
at some point of the inspiral (prior to the ISCO), where ${\hat
  L}_{\rm orb}$ is a unit vector parallel to the orbital angular
momentum.
\item Both the magnitude of the total spin ${\vec S_{tot}} = {\vec
  S_1} + {\vec S_2}$, and the angle $\theta_{\rm LS}$ between the total
  spin and the direction of the orbital angular momentum ${\hat
    L}_{\rm orb}$ will remain constant up to the ISCO.
\end{itemize}

We notice that in general $\theta_{\rm LS}$  and $| {\vec S_{\rm tot}} |$ can 
change during the evolution so this assumption might not hold to a tolerable
level. However, we know from the post-Newtonian spin-precession
equations~\cite{LK} that this assumption is valid in two cases, notably (i) 
equal-mass double-spin binary systems (when spin-spin terms are 
neglected) and (ii) unequal-mass single-spin binary systems~\cite{ACST}.

\begin{figure}
\begin{center}
\includegraphics[width=9cm,angle=-90]{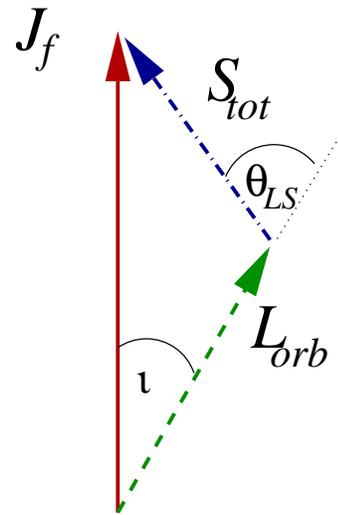}
\caption{\label{Fig:vectors} At ISCO, the total angular momentum will
  be the sum of the orbital angular momentum and the spin vectors.
  The angle $\beta$ is assumed to be given and remain fixed during the
  insprial prior to ISCO.  The inclination angle $\iota$ is solved for
  along with the magnitude of the total angular momentum.}
\end{center}
\end{figure}

From the initial conditions, we compute
\begin{eqnarray}
S_{\rm tot}&\equiv& | {\vec S_{\rm tot}} |, \\
\cos{\theta_{\rm LS}} &=& \frac{{\hat L}_{\rm orb} \cdot {\vec S_{\rm tot}}}{|\vec S_{\rm tot}|},
\end{eqnarray}
In Fig.~\ref{Fig:vectors}, we show the total spin and
orbital angular momentum at the ISCO, where ${\vec J_f} = 
{\vec L_{\rm orb}} + {\vec S_{\rm tot}}$.  As before we can then 
obtain the final spin of the black hole as $\vec a_f = {\vec J_f}/M$. 
More explicitly, if we decompose the vectors 
along the directions parallel and orthogonal to the 
final spin, we obtain
\begin{eqnarray}
L_{\rm orb}(\iota,a_f) \cos{\iota} + S_{\rm tot} \cos{(\theta_{\rm LS} - \iota)} &=& M a_f, 
\label{1} \\
L_{\rm orb}(\iota,a_f) \sin{\iota} - S_{\rm tot} \sin{(\theta_{\rm LS} - \iota)} &=& 0,
\label{2}
\end{eqnarray}
where $L_{\rm orb} = |\vec L_{\rm orb}|$. Eqs.~(\ref{1}-\ref{2}) can
be solved to derive the magnitude of the final spin $a_f$ and the
inclination angle $\iota$ at the ISCO.  For simplicity, we can compute
the orbital angular momentum of the inclined orbit using the fit
formula of Ref.~\cite{Hughes:2002ei}.  \bea
\label{eq:HBfit}
L_{\rm orb}(\iota,a_f) &=& \frac{1}{2} (1 + \cos{\iota}) L^{\rm pro}_{\rm orb}(r^{\rm pro}_{\rm ISCO},a_f) + \nonumber \\
&& \frac{1}{2}(1 - \cos{\iota})| L^{\rm ret}_{\rm orb}(r^{\rm ret}_{\rm ISCO},a_f)|, 
\eea
where $L_{\rm orb}^{\rm pro}$ and $L_{\rm orb}^{\rm ret}$ are given by Eq.~(\ref{eq:KerrL}) for
prograde and retrograde orbits, respectively, and $r^{\rm pro}_{\rm ISCO}$ and
$r^{\rm ret}_{\rm ISCO}$ are the corresponding ISCOs.  

With the above procedure, we can compare the obtained estimates with
available numerical results. In Table~\ref{tab:TichyMarronetti}, we
list the estimates from our approach together with the results
obtained from numerical simulations in Refs.~\cite{TM,loustofit,CLZM}
where several equal-mass double-spin and a few unequal-mass
single-spin precessing configurations have been studied. Note that in
cases ${\rm C}_1$ and ${\rm C}_3$, the total spin is zero (i.e. the
individual spins are equal, but opposite).  In any case where the
total spin is zero, our approach will predict the same result
(independent of the magnitude of the individual spins), which given as a
function of mass ratio by the bottom curve in
Fig.~\ref{Fig:spinsweep}.  Note that the cases ${\rm C}_6$, ${\rm
C}_7$ and ${\rm C}_8$ all have the same $S_{\rm tot}$ and $\theta_{\rm
LS}$, thus our approach predicts the same $a_f/M$ and $\iota$.  
Moreover, in cases ${\rm C}_1$ and ${\rm C}_3$, the masses are equal and 
the total spin is zero, but the individual spins are not 
on the orbital plane (corresponding to the large kick configurations in~\cite{CLZM,GSBHH07}). In these 
cases our approach predicts the same $a_f/M$ and $\iota$. Once
again in these more complex physical scenarios the agreement of our
simple approach is reasonably good.

\begin{table}
\begin{center}
\begin{tabular} {|c|c|c|c|c|} 
\hline
Case [Reference] & $a_f/M$ & $\iota$(deg) & $a_f/M$ & $\iota$(deg) \\\hline
 ${\rm C}_1$~\cite{TM} & $0.67$ & - &  $0.66$ & $0$ \\
 ${\rm C}_2$~\cite{TM} & $0.72$ & - & $0.71$ &  $23$\\
 ${\rm C}_3$~\cite{TM} & $0.68$ & - & $0.66$ & $0$ \\
 ${\rm C}_4$~\cite{TM} & $0.73$ & - & $0.71$ & $23$ \\
 ${\rm C}_5$~\cite{TM} & $0.64$ & - & $0.61$ & $34$ \\
 ${\rm C}_6$~\cite{TM} & $0.81$ & - & $0.82$ &  $14$\\
 ${\rm C}_7$~\cite{TM} & $0.80$ & - & $0.82$ & $14$ \\
 ${\rm C}_8$~\cite{TM} & $0.80$ & - & $0.82$ & $14$ \\
 ${\rm SP}_3$~\cite{loustofit} & $0.72 $ & 18 & $ 0.70$ & $21$  \\
 ${\rm SP}_4$~\cite{loustofit} & $0.81 $ & 10 & $ 0.80$ & $13$ \\
 ${\rm SP}_6$~\cite{CLZM} & $ 0.50 $ & $ 33 $ & $0.48$ & $35$ \\
 \hline
\end{tabular}
\end{center}
\caption{\label{tab:TichyMarronetti} In the second and third columns we 
list results from numerical simulations of precessing, spinning black holes, 
for the magnitude of the final spin and the 
angle between the final spin and the direction of 
the orbital angular momentum (if available). 
We use data from Ref.~\cite{TM} (we indicate with ${\rm C}_i, 
i = 1, ...8$ the spin configurations from left to right in Table I of Ref.~\cite{TM})
and from Refs.~\cite{loustofit,CLZM} (see Table I in each case). In the last two columns we list 
the estimates obtained with our approach.}
\end{table}

We find it important to stress again that it may be that for
more general precessional cases --- in which both 
black holes carry spin and their masses are not equal, or 
when the system undergoes a {\it transitional precession}~\cite{ACST} --- , 
the above approach may not provide a good approximation. 
We plan to investigate more generic cases in more detail in the future.

\section{Conclusions}
\label{sec:final-comments}

We have presented an approach to obtain a simple expression to
estimate the spin of the final black hole produced through a merger of
orbiting binary black holes. In this work we have concentrated on
several especially interesting cases, but others can certainly be
explored as well.

Notice that our work is complementary to recent works aimed at giving 
fit formulae for different physical quantities based on the results of
numerical simulations (see e.g., Refs.~\cite{loustofit,EOBp4PN,Rezzolla:2007xa,
Marronetti:2007wz,Boyle:2007sz}), and to the EOB
predictions~\cite{BD2,TD,BCD,DN}.  Our expression, however, does not
rely directly on the simulations, but rather on a simple approach
based on first principles. On the other hand, it has an inherent
amount of error due to its simple assumptions. Confronting our
predictions with available results, we find that they agree rather well
considering the limitations of our simplistic approach. This fact
gives further evidence to the rather simple behavior describing the
dynamics of orbiting black holes. The expression presented in this
work can be employed to predict the outcome of the simulation for a
large number of cases not yet studied and can help determine which
parameter choices might give the most interesting results.

For example:
\begin{itemize}
\item For individual spins aligned with the orbital angular momentum
  and $a_i \,\gaq\, 0.948 m_i$ the final black hole spin is larger as
  $\nu$ decreases. However for individual spins $a_i \,\laq\, 0.948
  m_i$, as $\nu$ is increased the final spin will be greater.  This
  {\it transition} number has an inherent error due to our
  approximations, and thus should not be considered sharp.
  Nevertheless, we would expect a transition to occur near this value.
  Hence our simple-minded model suggests that it is impossible to
  spin-up a black hole through mergers to its extremal value even in
  the {\it ideal} case where both spins and the orbital angular
  momentum are aligned, and no dissipative effects exist to reduce the
  final spin.  As indicated by Fig.~\ref{Fig:spinsweep}, the only way
  to get a final maximal spin requires an already maximally spinning
  black hole merging in extreme mass-ratio situations.  Any other
  alternative in the highly spinning cases ($a_i \,\gaq\, 0.948 m_i$)
  will cause the final spin to decrease. If ($a_i \,\laq\, 0.948 m_i$)
  the final spin will only increase up to a value $a \approx
  0.95$. After this state, any further merger will essentially leave
  this value unchanged\footnote{For a related discussion see
  Ref.~\cite{Hughes:2002ei}.}. Consequently binary black hole systems
  would not give rise to an orbital hang-up due to having $J > M^2$
  after the ISCO.
\item The direction of the orbital angular momentum of a system with
  arbitrary spins (perpendicular to the orbital plane) determines the
  final spin for $\nu \,\gaq\, 0.183$. For $\nu \,\laq\, 0.183$
  however the final spin direction will depend on how large the
  individual spins are; this can give rise to a final black hole whose
  spin opposes the initial orbital angular momentum direction. This
  scenario should give rise to an interesting phenomenology in the
  resulting waveforms.  We stress again that this critical value is
  approximate given our simple assumptions, but such a critical value
  must exist.
\end{itemize}

As discussed in Sec.~\ref{sec3C}, our approach can be generalized 
to spinning, precessing binaries. As long as the total spin of the system 
and the angle between the total spin and the orbital angular 
momentum are preserved during the evolution, we have been able to 
compare our results to numerical simulations of precessing 
binary systems, obtaining good agreement. We plan to carry out a more 
thorough study of more generic spinning, precessing binaries 
in the future and investigate further extensions of our approach.

Some other applications of this approach would be to employ the
predicted values in order to adopt a reasonably close background for
perturbative approaches to study the after-merger epoch or to compute
physical quantities with respect to this background like gravitational
radiation. Additionally it can be used to aid in providing the
quasi-normal mode frequencies (which are a function of the final black
hole mass and spin) to be used in analytically matching the inspiral
to the merger to the ringdown.

\acknowledgments
We would like to thank Latham Boyle, Michael Kesden, Samaya Nissanke
and William Unruh for interesting discussions, as well as Juhan Frank, Cole
Miller, Jorge Pullin and Saul Teukolsky  for useful comments. We would also
like to thank CIFAR and CITA for support and hospitality during the
Focus Group on Gravitational Waves and Numerical Relativity where this
work was started.

This work was supported in part by NSF grants PHY-0326311, PHY-0653369
and PHY-0653375 to Louisiana State University; NSF grant PHY-0603762
to the University of Maryland; and NSF grants PHY-0652952,
DMS-0553677, PHY-0652929, NASA grant NNG05GG51G, and a grant from the
Sherman Fairchild Foundation to Cornell University. A.B. also
acknowledges support from the Alfred P. Sloan Foundation.

\end{document}